\documentclass[conference]{IEEEtran}
\IEEEoverridecommandlockouts

\usepackage{cite}
\usepackage{amsmath,amssymb,amsfonts}
\usepackage{eso-pic}
\usepackage{graphicx}
\usepackage{textcomp}
\usepackage{xcolor}
\def\BibTeX{{\rm B\kern-.05em{\sc i\kern-.025em b}\kern-.08em
    T\kern-.1667em\lower.7ex\hbox{E}\kern-.125emX}}

\ifCLASSINFOpdf

\else

\fi
\usepackage{bm} 
\usepackage{siunitx}
\usepackage{times}  
\usepackage{helvet} 
\usepackage{courier}  
\usepackage[hyphens]{url}  
\usepackage{graphicx} 
\usepackage{graphicx}  
\usepackage[pagebackref=true,breaklinks=true,colorlinks,bookmarks=false]{hyperref}
\hypersetup{
    colorlinks=true,
    filecolor=magenta,      
    urlcolor=cyan,
}

\usepackage{subcaption}
\usepackage{amsmath,graphicx}

\usepackage{amsmath}
\usepackage{xcolor,soul,framed} 
\usepackage{xspace}
\usepackage{array}
\usepackage{eqparbox}
\usepackage{url}
\usepackage{longtable}
\usepackage{lipsum}
\usepackage{blindtext}
\usepackage{makecell}
\usepackage{mathtools}
\usepackage{commath}
\usepackage{multirow}
\usepackage{tabularx}
\usepackage[normalem]{ulem}
\usepackage{xspace}
\usepackage{algorithm}
\usepackage{algpseudocode}
\usepackage{amsfonts}
\usepackage{booktabs}
\usepackage{pifont}

\usepackage[table,xcdraw]{xcolor}

\newcolumntype{C}{>{\hsize=\dimexpr0.5\hsize+8\tabcolsep+\arrayrulewidth\centering\relax}X}

\DeclareMathOperator{\cref}{ref}

\newcommand{\cmark}{\ding{51}} 
\newcommand{\xmark}{\ding{55}} 

\begin{document}

\title{Development and Evaluation of Ultrasound Image Learning Pipelines for MASLD Risk Stratification\\
{\footnotesize }
\thanks{$\dagger$These authors contributed equally to this work. *Corresponding author: asamir@mgh.harvard.edu. This study was supported in part by NIH R01 DK\textcolor{black}{11}9860, the 2020 American Roentgen Ray Scholar Award, and 2019 Society of Abdominal Radiology Wylie J. Dodds Research award.}
}

\author{
\IEEEauthorblockN{
Guangyi Zhang\textsuperscript{1,$\dagger$},
Xiaohong Wang\textsuperscript{1,$\dagger$},
Eugene Cheah\textsuperscript{1},
Peng Guo\textsuperscript{1}}

\IEEEauthorblockN{
Brian A. Telfer\textsuperscript{2},
Theodore T. Pierce\textsuperscript{1},
Anthony E. Samir\textsuperscript{1, *}
}

\vspace{3mm}
\IEEEauthorblockA{\textsuperscript{1}
\textit{{Center for Ultrasound Research $\&$ Translation}, Massachusetts General Hospital, Harvard Medical School},
Boston, MA, USA
}
\IEEEauthorblockA{\textsuperscript{2}
\textit{Lincoln Laboratory, Massachusetts Institute of Technology}, Lexington, MA, USA
}
}

\AddToShipoutPictureFG*{%
  \AtPageUpperLeft{%
    \raisebox{-0.30in}{%
      \makebox[\paperwidth][c]{%
        \footnotesize
        Accepted and presented at IEEE EMBC 2026.
        \textcopyright~2026 IEEE.
      }%
    }%
  }%
}

\AddToShipoutPictureFG*{%
  \AtPageLowerLeft{%
    \raisebox{0.35in}{%
      \makebox[\paperwidth][c]{%
        \parbox{0.8\paperwidth}{%
          \centering
          \fontsize{8}{9}\selectfont
          \textcopyright~2026 IEEE. Personal use of this material is permitted.
          Permission from IEEE must be obtained for all other uses,
          in any current or future media, including reprinting/republishing
          this material for advertising or promotional purposes, creating
          new collective works, for resale or redistribution to servers or
          lists, or reuse of any copyrighted component of this work in
          other works.
        }%
      }%
    }%
  }%
}
\maketitle

\begin{abstract}
Metabolic dysfunction-associated steatotic liver disease (MASLD) affects approximately 30\% of the general population. Ultrasound-based imaging, including B-mode imaging and shear wave elastography (SWE), is widely used for noninvasive fibrosis assessment; however, the role of deep learning–based ultrasound image learning for MASLD risk stratification remains insufficiently characterized. 
In this study, we developed and evaluated ultrasound image learning pipelines using B-mode and SWE images for fibrosis staging and identification of patients with at-risk metabolic dysfunction-associated steatohepatitis (MASH). A total of 250 ultrasound examinations, one exam per subject, were included. Model performance was evaluated using 3-fold cross-validation with area under the receiver operating characteristic curve (AUROC). 
End-to-end SWE image learning achieved performance comparable to operator-guided SWE across fibrosis stages. Overall, SWE-based learning consistently outperformed B-mode image learning in fibrosis staging, with AUROC improvements from 0.64 (95\%CI: [0.56, 0.72]) to 0.72 (95\% CI: [0.65, 0.79]) for F$\geq$2 (\textcolor{black}{significant fibrosis,} p=0.11), from 0.67 (95\%CI: [0.58, 0.75]) to 0.78 (95\% CI:[0.72, 0.85]) for F$\geq$3 (\textcolor{black}{advanced fibrosis,} p=0.02), and from 0.69 (95\%CI: [0.56, 0.82]) to 0.80 (95\%CI: [0.72, 0.89]) for F4 (\textcolor{black}{cirrhosis}, p=0.10). These findings highlight the potential of SWE image learning for MASLD risk stratification.
\end{abstract}


\section{Introduction}
Metabolic dysfunction-associated steatotic liver disease (MASLD) is defined by hepatic steatosis with at least one cardiometabolic risk factor \cite{rinella2023aasld}. Its prevalence has increased rapidly over recent decades and currently affects approximately one-third of the U.S. population \cite{le2025estimated}. MASLD encompass a broad disease spectrum ranging from simple steatosis to metabolic dysfunction-associated steatohepatitis (MASH), fibrosis, and cirrhosis. A subset of patients with MASH may progress to hepatocelluar carcinoma, contributing substantially to morbidity, mortality, and economic burden \cite{younossi2025projected}. Among these patients, liver fibrosis is a major predictor of adverse clinical outcomes. Early detection and assessment on the degree of liver fibrosis is critical, as timely interventions can prevent severe liver-related outcomes\cite{younossi2025predictors}. 

Liver biopsy serves as the reference standard for fibrosis diagnosis and staging. However, the broad clinical application of liver biopsy on MASLD screening has constrained by its invasiveness, high cost, observer dependence, and procedure risks\cite{chowdhury2023liver}. There is a clear need for noninvasive, reliable, and scalable tool for fibrosis assessment. 

Conventional B-mode ultrasound (US) is widely due to its safety, accessibility, and low cost, enabling assessment of liver morphology changes associated with disease progression. while the interpretation of sonographic features is subjective\cite{lurie2015non}, limiting sensitivity for early-stage fibrosis detection. 

\textcolor{black}{US} elastography, including two-dimensional (2-D) shear wave elastography (SWE), provides quantitative liver stiffness measurement based on Young's modulus (kPa) or shear wave velocity (m/s) and has gained adoption for MASLD risk stratification. Compared with conventional B-mode, SWE offers greater objectivity, though its performance may be impacted by technical factors, patient habitus, and biologic confounders \cite{tang2015ultrasound}. Furthermore, the current clinical workflow for 2-D SWE imaging on certain vendor platforms requires manual placement of a circular region of interest (ROI) within the elasticity box. This is a time-consuming and operator-dependent step that is a barrier to effective population level screening.

To address these limitations, deep learning (DL), particularly convolutional neural networks, have been explored for prediction of liver fibrosis using \textcolor{black}{US} images \cite{huang2022evaluation, liu2023automatic, bose2025deep, punn2024liver, yousefzamani2025deep}. These approaches aim to capture subtle texture or patterns that may not be distinguished by human, potentially enabling improved detection of early fibrosis. Despite promising results, variability in data sets, model architectures, and evaluation protocols has constrained direct comparison with established methods.

\begin{figure}[htp!]
    \centering
    \includegraphics[width=0.9\columnwidth]{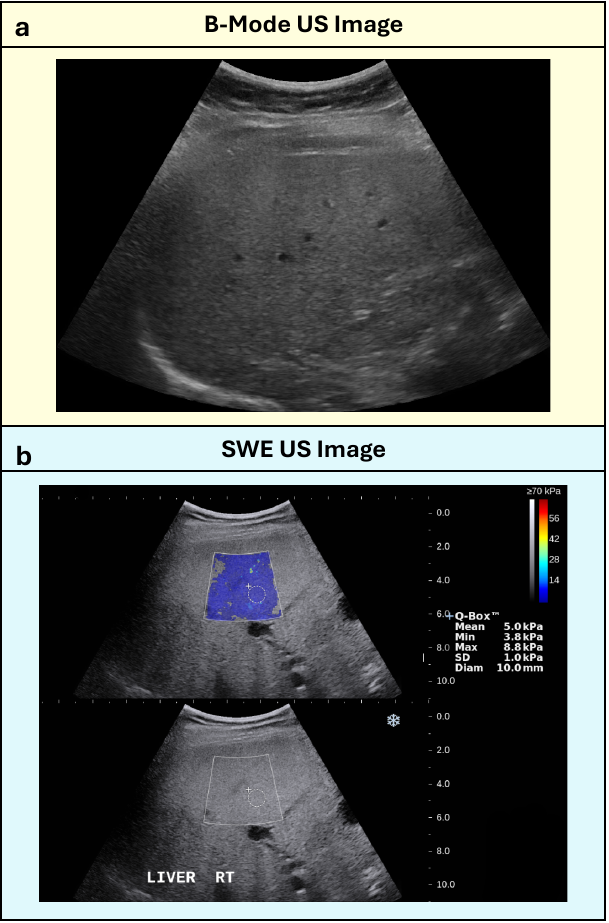}
    \caption{\textbf{Schematic of data modalities and representations.}
    \textbf{a}, Conventional B-mode ultrasound. Grayscale liver ultrasound images capturing parenchymal echotexture and anatomical structures were used as imaging inputs.
    \textbf{b}, SWE ultrasound. the elastography map is shown as a rectangular acquisition box overlaid on the B-mode image. Within this SWE box, a circular ROI is placed to sample liver stiffness measurements.
    }
    \label{overview}
\end{figure}

\textcolor{black}{In this work, we} develop and evaluate DL–based US image learning pipelines \textcolor{black}{using both B-mode and SWE images}, and compare these methods within a unified framework, \textcolor{black}{including operator-guided SWE as a reference. This enables systematic assessment of SWE image learning for fibrosis staging and at-risk MASH identification}. Our work seeks to lay a foundation for an end-to-end noninvasive screening tool to stratify the risk of liver fibrosis in MASLD patients.

\section{Study Population and Data Curation}
\subsection{Study Design}
This is a single-center, retrospective study approved by \textcolor{black}{the Mass General Brigham Institutional Review Board} (protocol no. 2021P003339). The requirement for informed consent was waived. 

\subsection{Study Population}
We identified adult participants who underwent non-focal liver biopsy from November 2013 to September 2021 at our institution, with 2-D SWE performed within one year. Exclusion criteria were: (a) absence or insufficient pathology information to determine fibrosis stage; (b) liver disease other than \textcolor{black}{clinically} confirmed MASLD/MASH; (c) liver stiffness $>$ 50 kPa; (d) SWE performed using devices other than the SuperSonic Imaging system (SSI Aixplorer); (e) absence of B-mode \textcolor{black}{US} images, as well as SWE images and extractable proprietary metadata from DICOM files; (g) SWE image dimensions other than 72L $\times$ 68W, to preserve the physical meaning of pixel values (liver stiffness) and avoid interpolation artifacts; (f) absence of demographic data and liver function test results within one year of liver biopsy.

As a result, the final data set comprising 250 participants (mean age $53 \pm 13$ years; $52\%$ women) were identified. The preliminary dataset included $2,552$ SWE images from $252$ \textcolor{black}{US} exams. To ensure one exam per participant for biopsy-SWE pairing, two exams were excluded, yielding $2,533$ SWE images from $250$ exams.

\subsection{US Procedure}
B-mode and shear wave elastography were performed using an Aixplorer scanner (SuperSonic Imagine, France) equipped with a curvilinear probe ($1$-$6$ MHz). Conventional B-mode US imaging was first used to obtain clear anatomical images, after which the SWE mode was activated to generate color-coded elasticity maps. SWE acquisitions were performed by trained sonographers through an intercostal window during neutral breath-holding, while avoiding major vessels and rib shadowing. A circular region of interest (ROI) was then placed within a stable and homogeneous elasticity area. Liver stiffness values (kPa) were automatically generated by the imaging system. \textcolor{black}{An example of B-mode and SWE liver US images used in this study is shown in Figure \ref{overview}.} The procedure was repeated ten times, and the median value of the $10$ measurements was used as the representative liver stiffness. 

\subsection{Pathology Reference Standard}
Liver biopsy procedures were conducted as per the standard protocol at our institution. We re-assessed nonfocal liver biopsy samples to ensure complete histopathologic evaluation, reduce inter-observer variability. 
Pathology slides were re-reviewed by one of three subspecialty pathologists with expertise in liver histopathology at our institution. Fibrosis staging was assessed using NASH Clinical Research Network (CRN) system \cite{sanyal2021prospective, sohal2025association}, which classifies fibrosis from stage F$0$ to F$4$ (no fibrosis, mild fibrosis, moderate fibrosis, bridging fibrosis, and cirrhosis).

\section{Prediction Tasks and Endpoints}
Subjects were stratified by fibrosis severity into four prediction targets: significant fibrosis (F$\geq2$), advanced fibrosis (F$\geq$3), cirrhosis (F$=4$), and at-risk MASH, defined as a MASH activity score $\geq4$ with fibrosis stages F$2$–F$3$. The prevalence of each target is summarized in Table~\ref{tab:prevalence} for the study cohort of $250$ participants. Prediction performance was evaluated using the area under the receiver operating characteristic curve (AUROC). Statistical comparisons between AUROCs were performed using DeLong’s test, with a p-value $< 0.05$ considered statistically significant. 

\begin{table}[!ht]
\centering
\caption{Prevalence of Prediction Targets in the Cohort.}
\label{tab:prevalence}
\small
\begin{tabular}{ccccc}
\toprule
\textbf{} & 
$\boldsymbol{\textbf{F}\geq2}$ & 
$\boldsymbol{\textbf{F}\geq3}$ & 
$\boldsymbol{\textbf{F}4}$ &
\textbf{At-Risk MASH} \\
\midrule
Cases & 71 (28\%) & 39 (16\%) & 14 (6\%) & 41 (16\%) \\
\bottomrule
\end{tabular}
\end{table}

\section{Data Modalities, Representations, and Learning Pipelines}

\begin{figure}[h!]
    \begin{center}
    \includegraphics[width=1.0\columnwidth]{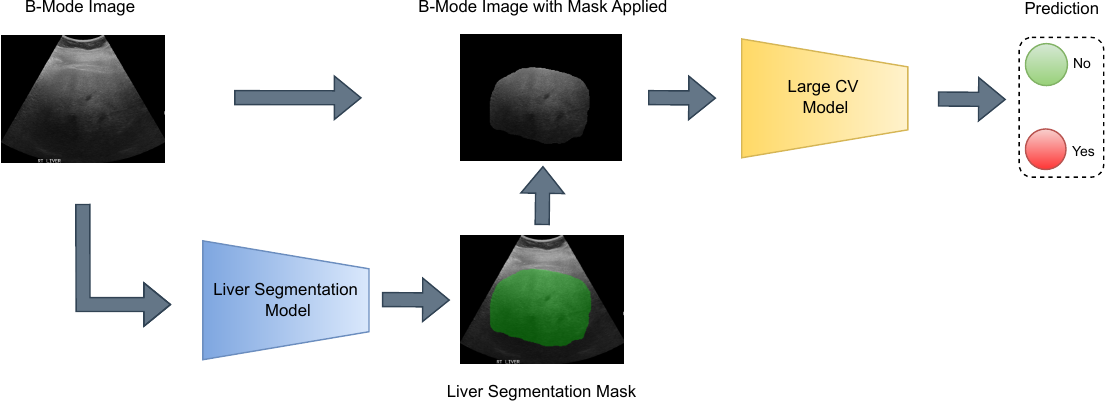} 
    \caption{B-mode learning pipeline. B-mode images containing liver tissue are identified using a pre-trained liver segmentation model \cite{ali2023liver}, and liver masks (shown in green) are applied to extract liver regions.}
    \label{b-mode}
    \end{center}
\end{figure}

\subsection{B-Mode Ultrasound}
As shown in Figure~\ref{b-mode}, we present a B-mode image learning pipeline for fibrosis staging and at-risk MASH prediction using B-mode \textcolor{black}{US} images as input. A total of $1,809$ B-mode images were available, of which $1,599$ were identified as containing liver tissue using a pre-trained liver segmentation model \cite{ali2023liver}. The relatively high spatial resolution of B-mode \textcolor{black}{US} images makes them well suited for established large-scale computer vision architectures.

Inspired by prior success in applying \textcolor{black}{DL} to liver fibrosis prediction from B-mode \textcolor{black}{US} images \cite{lee2020deep}, we evaluated a wide range of representative computer vision model families for B-mode image learning. These included ResNet \cite{he2016deep}, EfficientNet \cite{tan2019efficientnet}, ConvNeXt \cite{liu2022convnet}, Vision Transformer (ViT) \cite{dosovitskiy2020image}, BEiT \cite{bao2021beit}, Swin Transformer \cite{liu2021swin}, MaxViT \cite{tu2022maxvit}, and MambaOut \cite{yu2025mambaout}. For each model family, multiple architectural variants were considered (e.g., ResNet-$50$ and ResNet-$101$).

Models were evaluated both with and without ImageNet pre-trained weights. For models trained from scratch, images were normalized using the mean and standard deviation of the training set. For models initialized with ImageNet pre-trained weights, fine-tuning was performed using the standard ImageNet normalization. In addition, to study the impact of liver segmentation \cite{ali2023liver}, B-mode models were evaluated both with and without the application of segmentation masks (Figure~\ref{b-mode}).

Given the limited number of available B-mode \textcolor{black}{US} images, data augmentation was applied to expand the effective training set size by $10$-fold, in order to improve model robustness and mitigate overfitting. To reflect realistic \textcolor{black}{US} acquisition variability, conservative geometric transformations were applied during training, including random horizontal flipping and mild affine perturbations with constrained scaling ($0.95$–$1.05$), translation (up to $2\%$), and rotation ($\pm \ang{7}$). Intensity-based augmentations were carefully designed for B-mode \textcolor{black}{US}, incorporating limited brightness and contrast adjustments ($\pm 10\%$), additive Gaussian noise to approximate acquisition noise, and occasional mild Gaussian blurring to account for minor focus and resolution variations. 

\begin{figure*}[h!]
    \begin{center}
    \includegraphics[width=1.6\columnwidth]{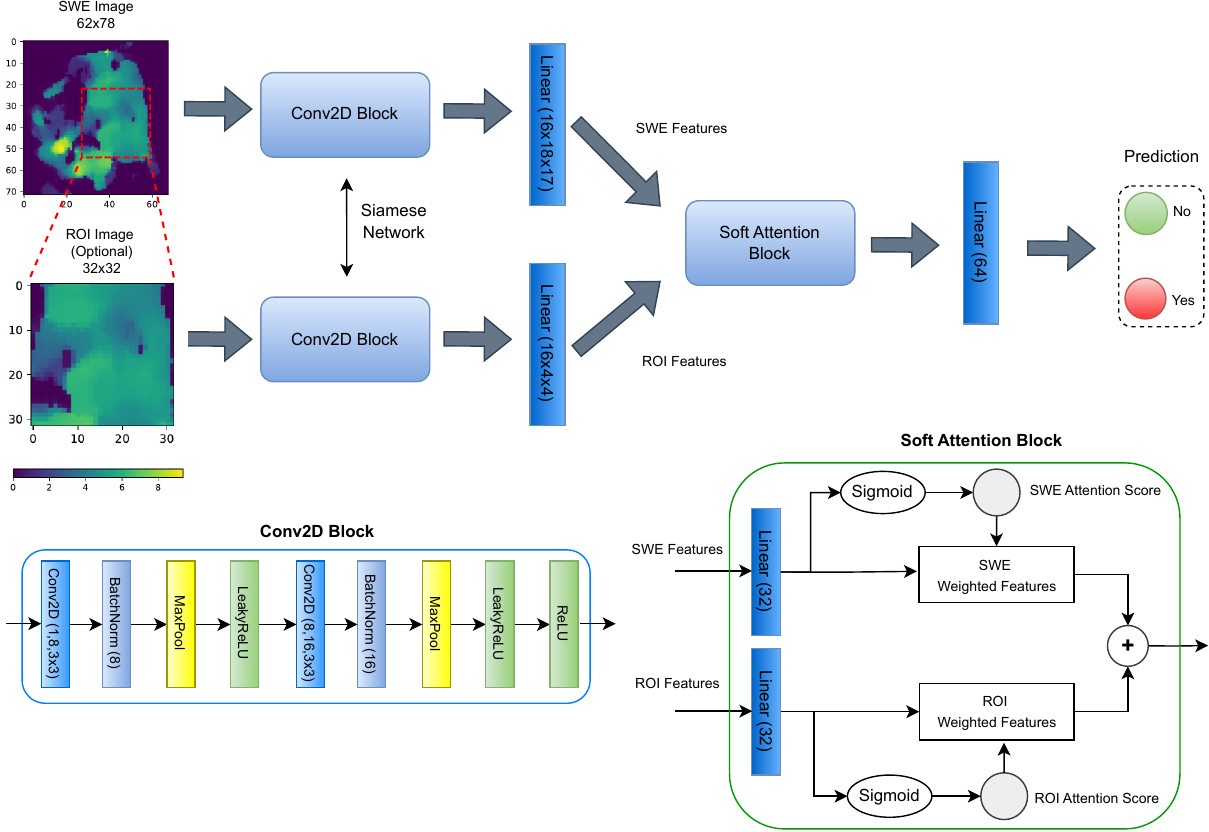} 
        \caption{Proposed Siamese network for paired SWE and ROI image learning. The architecture consists of two parallel Conv2D branches for SWE and ROI inputs, followed by a Soft Attention Block for adaptive feature fusion. Four operational modes are supported: SWE-Only, ROI-First, Single Attention, and Dual Attention. Numbers in brackets denote the number of neurons in linear or BatchNorm layers, while Conv2D annotations indicate input channels, output channels, and kernel size.}
    \label{swe-roi}
    \end{center}
\end{figure*}

\subsection{SWE Image}
\subsubsection{SWE Quality Assessment and ROI Extraction}
To address the varying quality of SWE images, which can hinder \textcolor{black}{DL} models from effectively capturing the discriminative fibrosis-related features, we adopted a two-step ROI extraction strategy, inspired by prior work \cite{brattain2020image}. 
First, SWE image quality was assessed using the percentage of color fill-in (PCFI), defined as the proportion of valid pixels within the elastography map. Only images with PCFI greater than or equal to a predefined threshold are retained for further analysis. Consistent with \cite{brattain2020image}, a threshold of $55\%$ was used, yielding $1,741$ qualified images from a total of $2,533$ SWE acquisitions.
Second, automated ROI extraction was applied to each qualified SWE image ($72\times68$). A $32\times32$ pixel scanning box slided across the image with a stride of one pixel to generate candidate ROIs. The ROI with the lowest standard deviation was selected to ensure a homogeneous region, reducing artifacts such as blood vessels and lesions \cite{brattain2018objective}. This procedure effectively mitigates noise and enhances the reliability of extracted ROIs for downstream fibrosis assessment. 
Prior to model input, SWE images and extracted ROIs were normalized using standard z-score normalization.

\subsubsection{Model Details}
We developed a Siamese network for paired SWE and ROI image learning to support fibrosis staging and at-risk MASH prediction.
As shown in Figure \ref{swe-roi}, the network comprised two parallel convolutional $2$D (Conv$2$D) blocks that independently process SWE and ROI images. The two blocks shared the same architecture design but do not share the trainable parameters. Each Conv$2$D block included two sub-networks, each composed of a convolutional $2$D layer, followed by batch normalization, max pooling, and LeakyReLU activation.
Features extracted from the SWE and ROI images were transformed via linear layers to ensure compatibility in feature dimensions and were subsequently passed to a Soft Attention Block for dynamic feature fusion.
Attention scores were computed using a sigmoid activation to quantify the relative contribution of SWE and ROI features. These scores were applied to weight the corresponding feature representations, which were then concatenated and passed to a fully connected layer for final prediction. An ROI availability flag was incorporated to indicate whether ROI inputs were valid or zero-padded. The network weights were initialized using Kaiming initialization \cite{he2015delving} to improve the training stability and convergence.

To accommodate variable ROI availability and to evaluate the contribution of ROI information, four operational modes were defined: SWE-Only, ROI-First, Single Attention, and Dual Attention. In SWE-Only mode, only SWE features were used, with ROI attention weights set to zero. ROI-First mode prioritized ROI features when valid ROI data were available, masking SWE features accordingly. Single Attention mode applied attention to both SWE and ROI features while using the ROI availability flag to gate ROI contributions. Dual Attention mode jointly learned attention weights for SWE and ROI features without explicit masking, allowing both modalities to contribute during training. Together, these modes enable flexible and robust modeling across varying ROI availability scenarios.

\section{Training and Evaluation Protocol}
The SWE image learning model was trained for $10$ epochs with a batch size of $64$ using the Adam optimizer and a learning rate of $0.001$. The B-mode image learning models were trained for $20$ epochs with a batch size of $128$ using the AdamW optimizer and a learning rate of $10^{-4}$. All experiments employed $3$-fold subject-level cross-validation with a fixed random seed, ensuring that images from the same patient were never shared between the training and testing sets within each fold, and were conducted in PyTorch $2.0$ on an NVIDIA A$100$ GPU. No hyperparameter tuning was performed. 
\textcolor{black}{Predictions from all held-out folds were subsequently aggregated, and performance metrics were calculated on the combined out-of-fold prediction set rather than averaged across folds.}

\section{Results}

\subsection{Performance of B-mode Ultrasound Image Learning}
Table \ref{tab:bmode_results} reports the prediction performance of representative B-mode \textcolor{black}{US} models across architecture families and clinical endpoints. For each family, we present the best-performing variant selected based on mean AUROC across all tasks, evaluated under four configurations defined by the presence or absence of ImageNet pretraining and liver segmentation guidance. Results are reported for significant fibrosis (F$\geq2$), advanced fibrosis (F$\geq3$), cirrhosis (F$=4$), and at-risk MASH. 

\begin{table*}[!h]
    \centering
    \caption{\textcolor{black}{AUROC of Best-Performing B-mode Ultrasound Models for Liver Fibrosis Prediction Across Architecture Families}}
    \renewcommand{\arraystretch}{1.1}
    \small
    \begin{tabular}{llccccccc}
    \toprule
    \textbf{Family} & \textbf{Architecture} & \textbf{Input} & \textbf{Seg.} & \textbf{Pretrain} & $\boldsymbol{\textbf{F}\geq2}$ & $\boldsymbol{\textbf{F}\geq3}$& \boldsymbol{$\textbf{F}4$} & \textbf{At-Risk MASH} \\
    \midrule
    \multirow{4}{*}{ResNet \cite{he2016deep}} & \multirow{4}{*}{ResNet-50} & \multirow{4}{*}{224}
    & \xmark & \xmark & 0.41 & 0.41 & 0.61 & 0.48 \\
    & & & \xmark & \cmark & $\mathbf{0.64}$ & 0.55 & 0.62 & $\mathbf{0.61}$ \\
    & & & \cmark & \xmark & 0.45 & 0.45 & 0.62 & 0.45 \\
    & & & \cmark & \cmark & 0.63 & $\mathbf{0.67}$ & $\mathbf{0.69}$ & 0.52 \\
    \midrule
    \multirow{4}{*}{EfficientNet \cite{tan2019efficientnet}} & \multirow{4}{*}{EfficientNet-B3} & \multirow{4}{*}{300}
    & \xmark & \xmark & 0.40 & 0.46 & 0.64 & 0.51 \\
    & & & \xmark & \cmark & 0.63 & 0.53 & 0.65 & 0.58 \\
    & & & \cmark & \xmark & 0.52 & 0.46 & 0.56 & 0.47 \\
    & & & \cmark & \cmark & 0.62 & 0.59 & 0.66 & 0.53 \\
    \midrule
    \multirow{4}{*}{ConvNeXt \cite{liu2022convnet}} & \multirow{4}{*}{ConvNeXt-B} & \multirow{4}{*}{224}
    & \xmark & \xmark & 0.45 & 0.48 & 0.62 & 0.41 \\
    & & & \xmark & \cmark & 0.59 & 0.53 & 0.62 & 0.56 \\
    & & & \cmark & \xmark & 0.51 & 0.48 & 0.55 & 0.54 \\
    & & & \cmark & \cmark & 0.58 & 0.52 & 0.60 & 0.55 \\
    \midrule
    \multirow{4}{*}{ViT \cite{dosovitskiy2020image}} & \multirow{4}{*}{ViT-B/16-384} & \multirow{4}{*}{384}
    & \xmark & \xmark & 0.53 & 0.55 & 0.67 & 0.61 \\
    & & & \xmark & \cmark & 0.54 & 0.53 & 0.61 & 0.52 \\
    & & & \cmark & \xmark & 0.59 & 0.57 & 0.55 & 0.55 \\
    & & & \cmark & \cmark & 0.53 & 0.61 & 0.65 & 0.51 \\
    \midrule
    \multirow{4}{*}{BEiT \cite{bao2021beit}} & \multirow{4}{*}{BEiT-B/16} & \multirow{4}{*}{224}
    & \xmark & \xmark & 0.48 & 0.55 & 0.66 & 0.54 \\
    & & & \xmark & \cmark & 0.55 & 0.49 & 0.58 & 0.58 \\
    & & & \cmark & \xmark & 0.50 & 0.53 & 0.58 & 0.53 \\
    & & & \cmark & \cmark & 0.48 & 0.48 & 0.58 & 0.55 \\
    \midrule
    \multirow{4}{*}{Swin \cite{liu2021swin}} & \multirow{4}{*}{Swin-T} & \multirow{4}{*}{224}
    & \xmark & \xmark & 0.38 & 0.44 & 0.60 & 0.55 \\
    & & & \xmark & \cmark & 0.57 & 0.48 & 0.64 & 0.57 \\
    & & & \cmark & \xmark & 0.47 & 0.46 & 0.52 & 0.50 \\
    & & & \cmark & \cmark & 0.54 & 0.53 & 0.62 & 0.55 \\
    \midrule
    \multirow{4}{*}{MaxViT \cite{tu2022maxvit}} & \multirow{4}{*}{MaxViT-S} & \multirow{4}{*}{224}
    & \xmark & \xmark & 0.47 & 0.51 & 0.56 & 0.46 \\
    & & & \xmark & \cmark & 0.56 & 0.55 & 0.62 & 0.60 \\
    & & & \cmark & \xmark & 0.45 & 0.48 & 0.60 & 0.51 \\
    & & & \cmark & \cmark & 0.54 & 0.60 & 0.62 & 0.57 \\
    \midrule
    \multirow{4}{*}{MambaOut \cite{yu2025mambaout}} & \multirow{4}{*}{MambaOut-S} & \multirow{4}{*}{224}
    & \xmark & \xmark & 0.50 & 0.50 & 0.55 & 0.53 \\
    & & & \xmark & \cmark & 0.58 & 0.51 & 0.64 & 0.65 \\
    & & & \cmark & \xmark & 0.44 & 0.44 & 0.57 & 0.49 \\
    & & & \cmark & \cmark & 0.58 & 0.62 & 0.54 & 0.59 \\
    \bottomrule
    \end{tabular}
    \vspace{1mm}
    \caption*{Note: backbones ($n=26$) included ResNet-34/50/101, EfficientNet-B0–B3, ConvNeXt-Tiny/Small/Base/Large; ViT-Small/Base/Large with 224 and 384 input, BEiT-Base with 224 and 384 input, Swin-Tiny/Small/Base, MaxViT-Tiny/Small; and MambaOut-Tiny/Small/Base. For each architecture family, we report the best-performing variant based on mean AUROC across all tasks. Input: image resolution in pixels (square); Seg.: liver segmentation mask; Pretrain: ImageNet pretrained weights.}
    \label{tab:bmode_results}
\end{table*}

\begin{table}[!htp]
\centering
\caption{\textcolor{black}{Impact of Pretraining and Segmentation Across All Evaluated B-mode Image Learning Architectures}}

\small
\begin{tabular}{ccccc}
\toprule
\textbf{Task} &
\multicolumn{2}{c}{\textbf{Pretraining}} &
\multicolumn{2}{c}{\textbf{Segmentation}} \\
 & Mean $\Delta$ & $p$-value &
   Mean $\Delta$ & $p$-value \\
\midrule
F$\geq2$
 & +0.09 & $<0.001$
 & +0.01 & 0.046 \\
F$\geq3$
 & +0.06 & $<0.001$
 & +0.02 & 0.094 \\
F$4$
 & +0.02 & 0.020
 & $-0.01$ & 0.446 \\
At-Risk MASH
 & +0.05  & $<0.001$
 & $-0.02$ & 0.059 \\

\bottomrule
\end{tabular}
\vspace{1mm}
\vspace{1mm}
\captionsetup{font=small}
\caption*{Note: paired effect-size analysis of pretraining and segmentation guidance across \textcolor{black}{all evaluated architecture variants described in Table II} ($n=26$), reported as mean $\Delta$AUROC (with minus without). Significance was assessed using the Wilcoxon signed-rank test. }

\label{tab:impact_b_mode}
\end{table}

\begin{table*}[!h]
\centering
\caption{AUROC for SWE Discrimination of Liver Fibrosis}
\label{tab:swe_results}
\small
\setlength{\tabcolsep}{6pt}
\begin{tabular}{lllcccc}
\toprule
\textbf{Method Type} &
\textbf{Method} &
\textbf{ROI Extraction} &
$\boldsymbol{\textbf{F}\geq2}$ & 
$\boldsymbol{\textbf{F}\geq3}$ & 
\boldsymbol{$\textbf{F}4$} &
\textbf{At-Risk MASH} \\
\midrule
Operator-guided SWE
& Median LS
& Sonographer-guided
& $0.70$ & $0.78$ & $\mathbf{0.86}$ & $\mathbf{0.63}$ \\
\midrule
Deep learning
& ROI-First
& Automatic
& $0.67$ & $0.77$ & $0.79$ & $0.55$ \\

Deep learning
& SWE-Only
& None
& $\mathbf{0.72}$ & $\mathbf{0.78}$ & $0.76$ & $0.59$ \\

Deep learning
& Single Attention
& Automatic
& $0.70$ & $0.77$ & $0.80$ & $0.59$ \\

Deep learning
& Dual Attention
& Automatic
& $0.72$ & $0.76$ & $0.76$ & $0.58$ \\

\bottomrule
\end{tabular}
\end{table*}

Table~\ref{tab:impact_b_mode} summarizes a paired effect-size analysis of pretraining and segmentation across all evaluated B-mode ultrasound models ($n=26$). Pretraining consistently improved performance across clinically relevant endpoints, with the largest gain observed for early fibrosis detection (F$\geq2$; mean $\Delta$AUROC = $+0.09$, $p < 0.001$), followed by F$\geq3$ ($+0.06$, $p < 0.001$) and at-risk MASH ($+0.05$, $p < 0.001$); gains for cirrhosis (F$4$) were modest ($+0.02$, $p = 0.020$). In contrast, segmentation guidance provided limited and task-dependent benefits, yielding a small improvement for F$\geq2$ ($+0.01$, $p = 0.046$) but no significant gains for F$\geq3$, F$4$, or at-risk MASH. Overall, pretraining emerged as the primary driver of performance improvement in B-mode \textcolor{black}{US}-based MASLD modeling, while segmentation had minimal impact.

\subsection{Performance of SWE-based Ultrasound Image Learning}
Of note, $56.4$ \% ($141$/$250$) of examinations in this study met criteria for high-quality (IQR/Median $\leq$ $30$\%) SWE images. The performance of SWE-based approaches, including the operator-guided, vendor-generated liver stiffness measure (using median value as representative) and four end-to-end DL methods \textcolor{black}{is} displayed in Table~\ref{tab:swe_results}. The operator-guided approach achieved the numerically highest AUROC values for predicting cirrhosis ($0.86$ [$95\%$CI: $0.71$, $0.85$]), and at-risk MASH ($0.63$ [$95\%$CI: $0.54$, $0.71$]). While the SWE-only DL method yielded the highest AUROC for significant fibrosis ($0.72$ [$95\%$CI: $0.66$, $0.79$]) and advanced fibrosis ($0.78$ [$95\%$CI: $0.66$, $0.79$]). Delong's tests were performed comparing each DL method with the operator-guided approach. No statistically significant differences were observed ($p>0.05$).

\subsection{Performance Comparison across B-mode learning, operator-guided SWE, and SWE Learning Approaches}
We compared the best-performing approaches for B-mode image learning, feature extraction based on operator-guided ROI placement in SWE, and SWE image learning.

For B-mode image learning, the ResNet model without segmentation and with pretrained weights was used for predicting F$\geq2$ and at-risk MASH, whereas the ResNet model with segmentation and pretrained weights was selected for predicting F$\geq3$ and F$4$. For SWE image learning approaches, the SWE-only \textcolor{black}{DL} model was used for predicting F$\geq2$ and F$\geq3$, while the Single-Attention \textcolor{black}{DL} model was selected for F$4$ and at-risk MASH.

\begin{figure}[ht!]
    \centering
    \includegraphics[width=1.0\columnwidth]{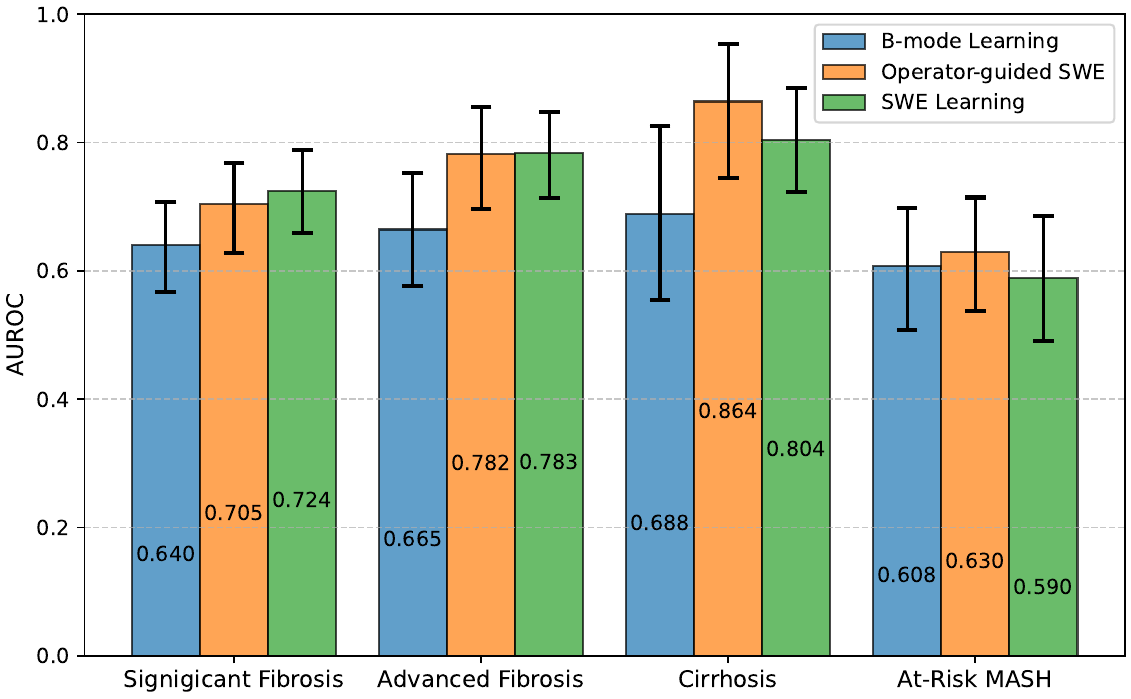}
    \caption{Performance comparison among B-mode learning, operator-guided SWE, as well as SWE learning approaches.}
    \label{comparison}
\end{figure}

Figure \ref{comparison} shows that, for significant fibrosis (F$\geq2$), our SWE learning approach achieved the numerically higher AUROC compared to operator-guided SWE approach ($0.72$ [$95\%$CI: $0.65$, $0.79$] vs. $0.71$ [$95\%$CI: $0.63$, $0.78$], $p=0.64$). For advanced fibrosis (F$\geq3$), both operator-guided SWE and SWE learning demonstrated significantly higher AUROC values than B-mode learning, (operator-guided SWE vs. B-mode learning: $0.78$ [$95\%$CI: $0.70$, $0.86$] vs. $0.67$ [$95\%$CI: $0.58$, $0.75$], $p=0.04$; SWE learning vs. B-mode learning: $0.78$ [$95\%$CI: $0.72$, $0.85$] vs. $0.67$ [$95\%$CI: $0.58$, $0.75$], $p=0.02$), while no statistically significant difference between operator-guided SWE and SWE learning ($p=0.98$). For cirrhosis, operator-guided SWE and SWE learning approaches show comparable AUROCs: $0.86$ [$95\%$CI: $0.76$, $0.97$] vs. $0.80$ [$95\%$CI: $0.72$, $0.89$], $p=0.33$). For the identification of at-risk MASH, the overall performance of the comparison methods underperformed compared to fibrosis staging prediction.

\section{Discussion}
In this study, we demonstrate the feasibility of using \textcolor{black}{DL}–based image learning approaches for MASLD risk stratification. The discrimination performance of different methods varies across fibrosis stages and at-risk MASH groups, which highlights the importance of task-specific method selection. Our major findings are: (1) End-to-end SWE learning shows comparable performance with operator-guided SWE across all the tasks, which has the potential to reduce time and costs when implemented into the clinical workflow and/or in the large population screening, particularly for F$\geq2$ and F$\geq3$; (2) In general, SWE learning outperforms B-mode learning.

Potential contributing factors to the suboptimal performance of all methods in at-risk MASH prediction include intrinsic signal-to-noise limitations in \textcolor{black}{US} imaging, heterogeneity in image acquisition, and weak alignment between local imaging features and composite underlying biologic activity including inflammation.

We also found that increased architectural complexity and larger parameter sizes (e.g., ResNet-$50$ vs. ViT-base-$384$) does not necessarily lead to improved predictive performance. Since B-mode and SWE image-based models may learn complementary information compared to operator-guided SWE measurement, we expect that fusing the three approaches will improve prediction accuracy. This will be the focus of future work. 

Despite the promising results, several limitations in this study should be noted. First, SWE image quality was limited, as shown by the high percentage of cases with elevated IQR/Median ratio. Potential causes include operator inexperience with the technique, insufficient time for exam completion as many were performed immediately prior to biopsy, or lack of clarity around imaging protocols. Local exam quality has significantly improved following rigorous quality improvement projects\cite{baikpour2025improving}. Second, \textcolor{black}{this single-center study has a modest sample size and a limited number of cirrhosis cases, which may affect generalizability. In addition, all \textcolor{black}{US} images were acquired using a single system, which may further limit the generalizability of the findings across different devices and vendors. }Although our SWE image learning approaches demonstrated performance comparable to the operator-guided method, \textcolor{black}{further investigation in larger, multicenter cohorts is needed to validate these findings. Third,} clinical impact and workflow integration remain to be established.

\section{Conclusion}
In this study, end-to-end SWE learning achieved performance comparable to operator-guided SWE across tasks indicating potential for reduced time and cost in clinical workflows and large-scale liver fibrosis screening in patients with MASLD.

\section*{Acknowledgment}
We thank Laura Brattain and Abder-Rahman Ali for brief discussions of the methods described in their papers.
We also thank the following researchers for their contributions to the MGH dataset during their time at the institution, which were incorporated into this study:
Arinc Ozturk, Marian Martin, Angela Shih, Atul \textcolor{black}{K. }Bhan, 
Joseph Misdraji, Firouzeh Heidari, Hannah Edenbaum, Madhangi (Maddy) Parameswaran, Katie Pope, Siddhi Hedge, Sai
Dhanush Reddy Jeggari, and Kim Naja.

\bibliographystyle{IEEEtran}
\bibliography{IEEEabrv,refs}

\vspace{12pt}

\end{document}